# ON THE LINK BEWEEN OCTAHEDRAL ROTATIONS AND CONDUCTIVITY IN THE DOMAIN WALLS OF BiFeO$_3$

Gustau Catalan[1,2]

[1]Institut Catala de Recerca i Estudis Avançats (ICREA), Barcelona, Spain
[2]Institut Catala de Nanociencia i Nanotecnologia (ICN2), Campus Universitat Autonoma de Barcelona, Bellaterra 08193, Spain.

## Abstract

We analyze the hypothetical link between octahedral straightening and increased conductivity inside the domain walls of BiFeO$_3$. Our calculations for 109° walls predict a lattice parameter expansion of c.a. 1% in the direction perpendicular to the wall, and an associated straightening of the octahedral rotation angle of 4$^o$, which is comparable to that observed in the high temperature metallic phase of BiFeO$_3$. On the other hand, in the closely related family of rare-earth orthoferrites, straighter octahedra do not correlate with increased bandgap, which suggests that the correlation between octahedral straightening and bandgap reduction in BiFeO$_3$ is perhaps fortuitous and not necessarily the cause of increased conductivity at the walls.

## 1. INTRODUCTION

When Jim Scott hired me as a postdoc in his group, apparently the thing that most interested him from my CV was not my research on ferroelectrics (which was hardly going to impress the so-called "pope of ferroelectrics" [1]), but my PhD work on the metal-insulator transition of NdNiO$_3$. Serendipitously -though serendipity seems to happen very often in Jim's life- this would later come in handy during the investigation of the metal-insulator transition of BiFeO$_3$ [2] and the transport properties of its domain walls [3]. In our review about this material [4] we described qualitatively our hypothesis that the metal-insulator transition in bulk BiFeO$_3$, and the increase in its domain wall conductivity may be at least partially related to a straightening of the octahedral rotation angles: the straighter bonds lead to an increased orbital overlap and a reduction in the bandgap, much as it happens also in the perovskite nickelates [5]. This idea resonated with us, as it drew on my background on nickelates, and also on one Jim's earliest breakthroughs – his 1968 discovery of the octahedral rotation modes in SrTiO$_3$ [6] – bringing that early work together with one of his latest achievements, the 2008 discovery of the metal-insulator transition in BiFeO$_3$ [2].

Would that life were so simple. The full story of the conductivity in the domain walls in BiFeO$_3$ and ferroics in general is of course more complex than that and is still unravelling [7, 8, 9, 10, 11]. There are many factors involved apart from (perhaps even instead of) octahedral rotations, including issues such as defects and electrostatic discontinuities. In the particular case of BiFeO$_3$, magnetic interactions have also been emphasized [11], and are thought by some to be the dominant factor: in particular Dieguez and Iñiguez [12], have performed ab-initio calculations that suggest that the key factor decreasing the local bandgap could be the defect-induced appearance of ferromagnetic pairs, with octahedral tilting playing a negligible role.

It is therefore my intention in this paper to scrutinize in more detail our initial hypothesis about the link between octahedral straightening and increased conductivity inside the domain walls of BiFeO$_3$, in order to establish whether the mechanism is quantitatively significant. The lattice expansion transverse to the domain walls of BiFeO$_3$, and the resulting steric straightening of the octahedral rotation are estimated. Comparison with data on the dependence of bandgap and octahedral rotation as a function of temperature in bulk BiFeO$_3$ suggests on one hand that the octahedral straightening in the wall is indeed comparable to that observed in the high conductivity regime of bulk BiFeO$_3$, and may therefore be sufficient to cause local increased conductivity inside the wall. On the other hand, a comparative examination of the closely related family of rare earth orthoferrites shows the opposite correlation, namely, compositions with straighter octahedral have bigger bandgaps. This suggests that the correlation between reduced bandgap and octahedral rotation in BiFeO$_3$ is not sufficient proof of a causal link, and that other factors need to be considered.

## 2. SPONTANEOUS STRAIN IN THE DOMAIN WALL

The ferroelectric polarization of BFO (rhombohedral space group R3c [13]) is directed along the diagonal of the cube, with identical projections over the three Cartesian coordinate axes; therefore, **P$_0$**=(Px, Py, Pz) with Px=Py=Pz=P$_0$/$\sqrt{3}$ ≡P. The spontaneous deformation of the unit cell along each direction is (Voigt compressed notation) $s_i = Q_{ij} P_j^2$. By symmetry, the Q$_{11}$=Q$_{22}$=Q$_{33}$ and Q$_{12}$=Q$_{13}$=Q$_{23}$. For the sake of simplifying the discussion we leave out the shear components (Q$_{i4}$ in Voigt notation) and focus on the cartesian components of the strain; the arguments are qualitatively the same, and quantitatively the solution will be in the right order of magnitude. Since the polarization is identical along the three cartesian axes (rhombohedral symmetry), the spontaneous strain is also identical:

$$s_x = s_y = s_z = 2/3 Q_{12} P_0^2 + 1/3 Q_{11} P_0^2. \tag{1}$$

This equation quantifies the stretching of the unit cell along the x, y and z directions caused by the spontaneous polarization. Conversely, the partial or total cancelling of the polarization inside the domain wall will necessarily cause a change in its internal spontaneous strain. The simplest wall to analyze is the 109$^o$ wall, as its locus is the (100) plane, which is perpendicular to the x-axis [14]. Conveniently, this is also the wall in which conductivity was first measured [3], although we must stress that conductivity exists also in 180 and 71 degree walls [8]. A 109$^o$ wall located in the (100) plane separates a domain with polarization along (111) from one with polarization along (1-1-1) [14]. The polarization normal normal to the domain wall ($P_x$) is, in theory, constant and thus its electrostrictive contribution does not disappear at the wall, although this is not strictly correct in practice, as a small polar discontinuity exists at the wall [3]. The other two components of the polarization are zero by symmetry, so the polarization at the centre of the wall is (P$_x$, 0, 0).

Physically, we can treat the wall as a vertical thin film of inherent spontaneous strain $s_x=Q_{11}P^2$, $s_y=0$, $s_z=0$, coherently clamped between a pair of domains with spontaneous strains given by (1). Because the domains are generally much broader than the walls, their spontaneous strain is not too affected by the wall, so they can be treated as "substrates" onto which the wall is clamped. Thus, in order to keep coherence in the domain wall plane, only the perpendicular component of the domain wall's spontaneous strain ($s_x$) is allowed to relax, while $s_y$ and $s_z$ of the wall are forced to adopt the same strain state as the domains, as discussed by Zhirnov [15]. The strain component normal to the wall experiences a Poisson's ratio deformation imposed by the difference between the in-plane spontaneous strain of the unit cells inside the wall, $s_y=s_z=0$, and the strain state of the adjacent domains, $s_y=s_z= 2Q_{12}P^2 +Q_{11}P^2$. Thus, the perpendicular strain at the wall is

$$s_x^{wall} = Q_{11}\frac{P_0^2}{3} - (s_y+s_z)\frac{\nu}{1-\nu} = Q_{11}\frac{P_0^2}{3} - \left(\frac{4}{3}Q_{12}P_0^2 + \frac{2}{3}Q_{11}P_0^2\right)\frac{\nu}{1-\nu} \qquad (2)$$

And therefore the relative change with respect to the spontaneous strain in the domain is

$$s_x^{wall} - s_x = -\frac{2P_0^2}{3}\left(Q_{12}\left(\frac{1+\nu}{1-\nu}\right) + Q_{11}\frac{\nu}{1-\nu}\right) \qquad (3)$$

Let us now put some numbers into the equations.

The spontaneous polarization of BiFeO$_3$ along the rhombohedral <111> direction is, approximately, $P_0=1C/m^2$ [16] The electrostrictive components of BiFeO$_3$ are not known, but the piezoelectric components have been calculated and found to be smaller than those of PbTiO$_3$, which is another perovskite ferroelectric with a polarizable lone pair in the A-site). In perovskite ferroelectrics (proper ferroelectric, improper ferroelastic), the piezoelectric constant is linearly proportional to the electrostrictive coefficient: $d=Q\varepsilon P^2$, where $\varepsilon$ is the dielectric constant. Since PbTiO$_3$ has similar values of polarization and dielectric constant to those of BiFeO$_3$, we can assume as a first order approximation that the ratio between the piezoelectric coefficients of BiFeO$_3$ and PbTiO$_3$ is the same as the ration between the electrostrictive constants of BiFeO$_3$ and PbTiO$_3$. The longitudinal piezoelectric constant is roughly 1/3 of that of PbTiO$_3$, and the transverse one is roughly 1/2 [17], the expected electrostrictive coefficients of BiFeO$_3$ are, roughly, $Q_{11}^{BFO} = \frac{1}{3}Q_{11}^{PTO} = 2.97\times10^{-2} m^2C^{-2}$ and

$$Q_{12}^{BFO} = \frac{1}{2}Q_{12}^{PTO} = -1.3\times10^{-2} m^4C^{-2}.$$

Where the electrostrictive coefficients of PbTiO$_3$ have been extracted from ref. [18]. A set of electrostrictive coefficients has also been reported by Zhang et al [19], and is $Q_{11}=0.032$ C$^{-2}$m$^{-4}$ and $Q_{12}=-0.016$C$^{-2}$m$^4$, which is in remarkably good agreement given our simple approach. As for the Poisson's ratio, we assume a typical value for perovskites, $\nu=0.3$. Putting these numbers in eq. 3 we obtain that the change in spontaneous strain perpendicular to the 109° wall is approximately $\Delta s_x=1 \times 10^{-2}$ We are not aware of any first principles calculations or structural studies with which to compare this result; clearly it would be desirable that they be performed. Notice also that the above calculations assume that the lattice is unclamped in the x-direction, an assumption which is valid for free standing single crystals, but which probably needs corrected for epitaxial thin films where there is also in-plane clamping.

### 3. OCTAHEDRAL ROTATION

An exact calculation of the octahedral rotations and Fe-O-Fe bond angle requires consideration of at least three different contributions:

1) The geometrical coupling between the unit cell shape and the three-dimensional steric rotation of the octahedron as a rigid unit. In free energy terms, this is reflected by the rotostrictive coupling between spontaneous strain and octahedral rotation.
2) The biquadratic coupling between polarization and octahedral rotation via the rotopolar coefficient [20]. The free energy incorporates a term $\eta P^2\phi^2$ that, being biquadratic, is always symmetry-allowed and always positive in order to ensure thermodynamic stability [21]. A direct consequence of this term is that polarization and rotation frustrate each other, so that where the polarization is cancelled (e.g., in the middle of the ferroelectric domain wall) the rotation angle will tend to increase.
3) Even in the absence of octahedral rotations, the ferroelectric off-centering of the central cation means that the Fe-O-Fe bond is not straight, as the Fe is no longer coplanar with the oxygen in the $FeO_2$ atomic planes [12].

The above contributions are all inter-linked, and several key coefficients, such as the rotostrictive and rotopolar coupling coefficients, are not precisely known, which precludes an exact solution to the problem at this stage. Nevertheless, the orders of magnitude involved can be estimated using geometry. The octahedral rotation arises from the fact that the size of the unit cell is too small to have a "straight" octahedron, and thus the octahedra have to buckle in order to fit inside; if the unit cell is expanded in the domain wall, the buckling angle will necessarily decrease (see figure 1).

Domain Wall

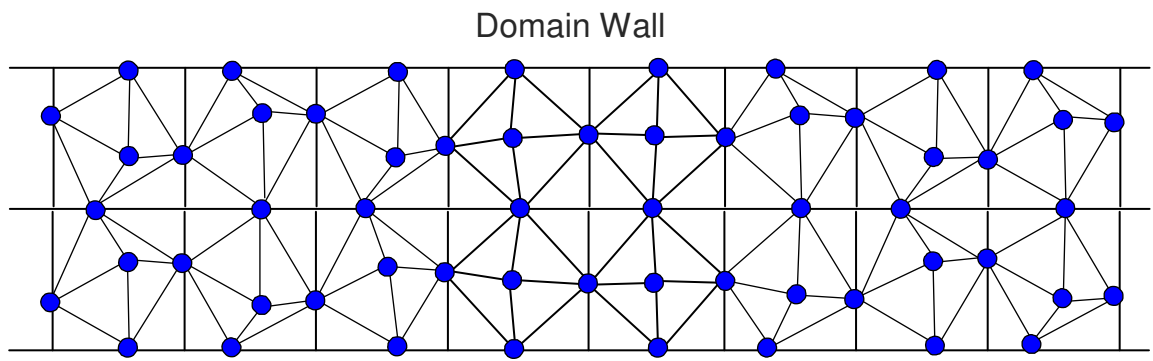

**Figure 1**. Octahedral tilts appear in order to fit the oxygen octahedral inside the perovskite pseudocubic unit cell; the unit cell expansion in the domain walls leads to a reduction of the octahedral buckling.

In perovskites, formula unit $ABO_3$, the buckling angle is directly related to the Goldschsmid's tolerance factor, defined as $t \equiv \dfrac{d_{A-O}}{\sqrt{2}d_{B-O}}$ (where $d_{A-O}$ is the distance between the A-site, bismuth in our case, and the oxygen, and $d_{B-O}$ is the distance between the central cation, iron in $BiFeO_3$, and the oxygen). When the tolerance factor is smaller than 1, rotations appear in order to fit the octahedral inside the pseudocube.

The rotation angle around the z-axis, $\phi_z$, can be related to $d_{O\text{-}Fe\text{-}O}$ and to the horizontal length of the perovskite unit cell, $a$ by

$$\cos\phi = \frac{a}{d_{O-Fe-O}} \quad (1)$$

From this it follows that any change in the unit cell size $a$ or in the $d_{M\text{-}O}$ bond length will affect the angle $\phi$. We note here that $\phi$ also affects the strength of the antiferromagnetic exchange interaction. As the horizontal lattice parameter of the domain wall is bigger (see the positive value of $s_x$ calculated earlier), it is expected that the tilting angle inside the domain wall will be bigger than outside. Quantitatively, we can use (1) to relate the change in tilting angle to the in-plane strain at the domain wall:

$$s = \frac{a-a_0}{a_0} = \frac{d_{O-Fe-O}\cos\phi}{d^0_{O-Fe-O}\cos\phi_0} - 1 \quad (2)$$

Where the index 0 refers to the unit cell outside of the domain wall. Thus,

$$\cos\phi = \cos\phi_0 \frac{d^0_{O-Fe-O}}{d_{O-Fe-O}}(s_x+1) \quad (3)$$

Let us put numbers into this. The known value for the octahedral rotation in bulk $BiFeO_3$ is $\omega=12$ degrees around the <111>, and the resolved rotation around the <001> axis, given by $\phi = \frac{\omega}{3^{1/3}}$ [20], is $\phi_0=8.3$ degrees around <001>. $s_x$ was estimated earlier ($s_x\approx 10^{-2}$). If the octahedra were rigid ($d^0_{Fe\text{-}O}/d_{Fe\text{-}O}=1$), the expansion of the lattice parameter would be more than enough to completely straighten the angle and in fact the octahedra would have to expand in order to fill the space. But the octahedra are NOT rigid, because their size and shape are affected by the ferroelectric off-centering of the Fe cation; since the polarization is at least partially suppressed inside the domain wall, the size of the octahedra must also be different. Arnold et al [22] have reported the exact change in the octahedra at the first order ferroelectric transition: there are two Fe-O distances in the ferroelectric phase; when added up, the O-Fe-O distance is 4.08Å. In the paraelectric phase, there is only one Fe-O distance, and the O-Fe-O is 4.06Å. The suppression of the polarization thus leads to a compressive octahedral strain $s_d = 5\times 10^{-3}$. If the polarization is only suppressed along the z and y directions, there will be a octahedral compression along the z and y axis that will result, via Poisson's deformation, in an expansion along the x direction; we can thus relate the distance O-Fe-O inside the wall can be related to that outside the wall by

$$\frac{d_{O-Fe-O}}{d^0_{O-Fe-O}} = \left(1 - s_d \frac{\nu}{1-\nu}\right) \cong 0.998 \quad (4)$$

When putting this value back into eq. 3, we obtain that $\phi\approx 4.1°$, that is, the octahedral rotation inside the wall is about half its value outside. It is interesting to note that the fact that $\frac{d^0_{Fe-O}}{d_{Fe-O}}<1$ partially offsets the positive contribution of $s_x$. Geometrically, then, the role of the polarization-induced change in octahedral size reflects the role of the rotopolar coupling: the bigger the polarization-induced narrowing of the octahedron, the smaller the octahedral tilting will be, reflecting the mutual frustration between P and $\phi$.

Note that the above calculations are only a first approximation that is necessarily inexact as we have neglected, among other things, the shear strain. We nevertheless take take 4 degrees as an indication of the order of magnitude of the octahedral straightening at the wall. The question is, then, what effect would such a change in rotation angle have on the local bandgap? For this, we turn to what is known the experimental data for bulk $BiFeO_3$.

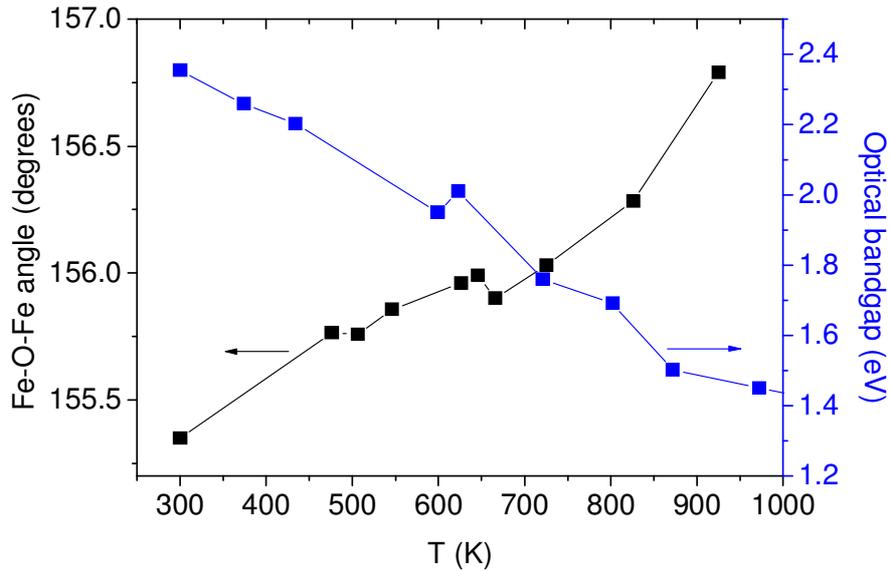

**Figure 2.** Evolution as a function of temperature of the Fe-O-Fe angle (black) and optical bandgap (blue) for bulk $BiFeO_3$. This graph was made by compiling the published data for bond angles from Palewicz et al [24] and the optical bandgap data reported by Palai et al [2].

In figure 2, first shown in the context of a discussion of the metal-insulator transition in $BiFeO_3$ [23], we overlapped the Fe-O-Fe bond angle as a function of temperature reported by Palewicz et al [24], with the optical bandgap as a function of temperature measured by Palai et al [2]. It is quite apparent that one is the mirror image of the other, including a possible anomaly around the Neel temperature (640K). Obviously correlation does not imply a causal link, and perhaps both optical bandgap and Fe-O-Fe angle are controlled by a third parameter that has not been taken into account. For example, the effect of spin ordering on bandgap has been emphasized [12], and indeed the activation energy for conduction above $T_{Neel}$ is known to be much larger than below it (see e.g. figure 7 in the review by Catalan&Scott [4]). Interestingly, there is no equivalent sudden drop of the optical bandgap at $T_{Neel}$, which continues to decrease well above the magnetic transition temperature; as Viret and Iñiguez argue [25], this probably reflects the fact that the bandgap is dictated by short range interactions rather than long range order.

At any rate, from figure 2 we can extract a phenomenological correlation between octahedral tilt and bandgap: the bandgap decreases roughly 1eV for each 1.5 degrees of bond straightening. Therefore, a straightening of c.a. 4 degrees would be, according to this model, sufficient to completely close the 2.5-2.8 eV bandgap [2, 26] and cause a transition to the metallic state. Since this critical straightening is within the expectation values calculated for the tilt straightening inside the wall, it would seem to validate our initial hypothesis for the link between conductivity and octahedral rotation. Comparison with other perovskite iron oxides, however, suggests that the decrease in bandgap inside the walls may not be due to octahedral straightening.

## 4. OCTAHEDRAL ROTATION IN ORTHOFERRITES

It has been pointed out before [27, 4] that $BiFeO_3$ is very close, magnetically and structurally to rare-earth othoferrites. In fact, above the ferroelectric Curie temperature, $BiFeO_3$ is indeed an orthoferrite, as it has exactly the same Pbnm orthorhombic space group [22, 29]. A comparison between the bandgaps of different rare-earth orthoferrites should therefore be illuminating. Notice that, in going from $LaFeO_3$ to $LuFeO_3$, the Fe-O-Fe bond angle decreases from 158 degrees to 141 degrees, and this reduction in the exchange angle leads to a significant reduction of the Neel temperature, from 730K to 620K [28]. The weakening of the magnetic interactions should therefore lead to a decreased bandgap in the magnetically-dominated bandgap scenario, whereas, conversely, increased tilting of the angle should lead to an increased bandgap in the octahedral rotation dominated scenario, so this comparison should provide a very useful tool to help decide.

Lyubutin *et al.* [30] report the optical bandgaps of $LuFeO_3$ and $NdFeO_3$, they are 2.25eV and 2.5eV respectively, even though the the Fe-O-Fe bond angles are 141° for $LuFeO_3$ and 151° for $NdFeO_3$ [28]. In other words, even though $NdFeO_3$ has the straighter bonds, it also has the bigger bandgap, suggesting that the bond angle is probably not the dominant effect the bandgap. Notice, however, that the comparison is nevertheless complicated by the fact that, aside from octahedral distortions and magnetic ordering, the inter-atomic distances are also different, being smaller in LuFeO3: perhaps the lattice contraction plays the dominant role in reducing the bandgap of orthoferrites. High pressure experiments [30] are consistent with this.

## CONCLUSION

In summary, we have examined the factors that determine the iron-oxygen-iron bond angle in the domain walls of $BiFeO_3$; the full calculation must incorporate the steric straightening that arises from the lattice expansion inside the wall, and the rotopolar coupling that arises from the direct coupling between polarization and rotation. These two terms have opposite effects. Using geometrical considerations, we have estimated that the tilting angle around the z axis decreases by about 4 degrees inside the domain wall. At the same time, the measured correlation between bond angle and optical bandgap in bulk $BiFeO_3$ suggests that a straightening of 4 degrees should be sufficient to close the bandgap completely.

On the other hand, analysis of the correlation between bond angle and bandgap in the closely related family of perovskite orthoferrites, shows the opposite trend. This suggests that the octahedral straightening inside the wall may affect other functional properties inside the wall [31] (notably the magnetism, which is very sensitive to superexchange angles), but is not necessarily the culprit, or at any rate not the only culprit, of the increased conductivity of the domain walls.